\documentclass[1, 12pt]{elsarticle}
\usepackage[english]{babel}
\usepackage{amsmath}
\usepackage[toc,page]{appendix}
\usepackage{amssymb}       
\usepackage{amsthm}       
\usepackage{natbib}        
\usepackage{graphicx}
\usepackage{graphics}
\usepackage{booktabs}
\usepackage{subfigure}      
\usepackage{color}
\usepackage{gensymb} 

\setcitestyle{square, comma, numbers,sort&compress}
\journal{Physical Chemistry Chemical Physics, accepted for publication}
\makeatletter
\def\@xfootnote[#1]{%
  \protected@xdef\@thefnmark{#1}%
  \@footnotemark\@footnotetext}
\makeatother

\begin{document}

\begin{frontmatter}

\title{The effect of elastic strains on the adsorption energy of H, O, and OH 
       in transition metals}

  \author[imdea,ucm]{Carmen Mart\'inez-Alonso}
  \author[upm]{Jos\'e Manuel Guevara-Vela}
  \author[imdea,upm]{Javier LLorca\corref{cor1}}
  \cortext[cor1]{To whom correspondence should be addressed: javier.llorca@upm.es , javier.llorca@imdea.org}

  \address[imdea]{IMDEA Materials Institute, C/Eric Kandel 2, 28906 - Getafe, Madrid, Spain.}
  \address[ucm]{Department of Inorganic Chemistry, Complutense University of Madrid, 28040 Madrid, Spain.}
  \address[upm]{Department of Materials Science, Polytechnic University of Madrid, E. T. S. de
Ingenieros de Caminos, 28040 Madrid, Spain.}

 \begin{abstract}

The influence of elastic strains on the adsorption of H, O, and OH on the (111)
surfaces of 8 fcc (Ni, Cu, Pd, Ag, Pt, Au, Rh, Ir) and on the (0001) surfaces
of 3 hcp (Co, Zn, Cd) transition metals was analyzed by means of density
functional theory calculations. To this end, surface slabs were subjected to
different strain states (uniaxial, biaxial, shear, and a combination of them) up
to strains dictated by the mechanical stability limits indicated by phonon
calculations. It was found that the adsorption energy followed the predictions
of the d-band theory but -surprisingly- 
the variations in the adsorption energy only depended on the area of the adsorption hole and not on the particular elastic strain tensor applied to achieve this area.  The analysis of the electronic structure
showed that the applied strains did not modify the shape of Projected Density
of State (PDOS) of the d-orbitals of the transition metals but only led to a
shift in the energy levels. Moreover, the presence of the adsorbates on the
surfaces led to negligible changes in the PDOS. Thus, the adsorption energies were
a function of the Fermi energy which in turn was associated to the change of
the area of the adsorption through a general linear law that was valid for all
metals. The information in this paper allows the immediate and accurate
estimation of the effect of any elastic strain on the adsorption energies of H,
O, and OH in 11 transition metals with more than half-filled d-orbitals.

 \end{abstract}


\end{frontmatter}

\section{Introduction}

The development of efficient and selective catalysts has been one of the
keystones of technological progress in the last century \cite{catalysis}.
Currently, most fossil fuels used in transportation are refined using catalytic
processes \cite{Lee2014,deLasa2011,Primo2014}, while a substantial fraction of
chemical products are manufactured through technologies based on catalysis
\cite{Jessop1994,Busca2007,Suryanto2019}. Additionally, catalysis is critical
for the electrochemical processes necessary to generate clean energy or/and to
produce clean fuels, such as hydrogen \cite{Norskov2006, Koper2010}. In
particular, hydrogen economy is based in two critical reactions that are
account for the production of hydrogen and the generation of clean energy,  the
hydrogen evolution reaction (HER) and the oxygen reduction reaction (ORR),
respectively \cite{Vesborg2015,Shao2016}. So far, Pt stands as the best
catalyst for these reactions but its high cost and limited availability
hinders the industrial application of this technology, leading to a search for
cheaper and more widely available catalysts \cite{Hansen2021}.

It is known that the rate limiting steps for both HER and ORR reactions are
associated with the adsorption of intermediate species (H, O, and OH) onto the
surface of the catalyst \cite{Nrskov2004,Dubouis2019}. These processes are controlled by
the electronic structure of the catalysts, that can be modified using different techniques, e.g. addition of alloying elements, introduction of defects and/or surface orientation \cite{EscuderoEscribano2016, Fu2019,ZamoraZeledn2021}.
Another effective mechanism to modify the electronic structure is through the
introduction of large elastic strains (1-3\%) in the lattice
\cite{Shuttleworth2016,Shuttleworth20177}. This strategy - denominated elastic
strain engineering - opens the possibility to modify the electronic, optical,
magnetic, and catalytic properties of solids through the systematic application
of elastic strains \cite{Feng2018, Wang2018,Kong2017,Rudi2012}.  Moreover,
recent experimental work has shown that much larger tensile elastic strains
(close to the theoretical limit of 10\%) can be attained in particular
nanomaterials (such as nanoneedle diamond structures as well as carbon
nanotubes) \cite{LLorca2018,Banerjee2018} and metallization of diamond has been
predicted at this strain level by means of first principles calculations
\cite{Shi2019}. It is obvious that dramatic changes in both the electronic
structure and the catalytic properties of materials could be expected when the
elastic strains approach the instability limit. However, this territory is
unexplored due to the experimental difficulties associated with the application
of such large elastic strains.  

The effect of mechanical strains on the electronic properties  of transition
metals, which are the most important for the HER and ORR reactions, has been
qualitatively analyzed by many authors
\cite{Bhattacharjee2016,Grunze1982,Rao1991,Ruban1997,Cheng1995,Xin2014}.  Their
findings can be summarized in the so-called  d-band center model
\citep{Hammer2000, Hammer1995, Hammeeer1995}, developed more than two decades
ago. This model is based on the effective medium theory
\cite{PhysRevB.35.7423,Jacobsen1996} and the Newns-Anderson model
\cite{NEWNS1969,Anderson1961}, and relates the adsorption energy to the change
in the local electron density of the surface. The changes in the adsorption
energy with mechanical strain in the d-band model are related to the change of
the energy of the d-band center, $\epsilon_{d}$, and the variation in the
adsorption energy from one transition metal surface to another correlates with
the upward shift of $\epsilon_{d}$ with respect to the Fermi level. A  shift
towards higher energies allows the formation of a larger number of empty
anti-bonding states, leading to a stronger (more negative) binding energy. Even
though the d-band model can be used to rationalize the influence of elastic strains in the catalytic activity, there is not a model to predict directly the adsorption energy as a function of the
applied strain using $\epsilon_{d}$.

In this investigation, the relationship between the applied elastic strain
tensor and the adsorption energy of H, O, and OH on the surface of eleven
transition metals is determined by means of density functional theory (DFT)
calculations. A simple relationship between the area of the deformed surface
hole where the adsorbates lay and the energy associated with the adsorption
process is found for all metals. It was also determined that the adsorption
energy only depends on the deformed area of the hole and is independent of the
elastic strain tensor applied to achieve this area. Thus, variations in the
adsorption energy with elastic strain could be obtained with very limited
computational effort. Moreover, a linear relationship between the adsorption
energy and the Fermi energy was also found for all metals, the
slope indicating the sensitivity of each metal to change the catalytic
properties through elastic strain engineering. This information is relevant as
a first approach to provide a quantitative understanding of the application of elastic
strain engineering to modulate the adsorption energy of transition metals and
support the search for better catalysts.

\section{Computational details}

The DFT plane wave simulations were performed using the Quantum Espresso
package \cite{Giannozzi2009}. The electron exchange-correlation was described
using the generalized gradient approximation (GGA) with the
Perdew-Burke-Ernzerhof (PBE) exchange-correlation functional \cite{Perdew1996}
and the calculations were carried out using ultrasoft pseudopotentials.
Brillouin zone calculations were performed using a
Marzari-Vanderbit-DeVita-Payne cold smearing of 0.015 Ry \cite{Marzari1999}.
The planewave basis was expanded to a cutoff energy of 80 Ry, and the
Monkhorst-Pack k-points were sized ($20 \times 20 \times 20$) for the unit
cells of all metals, and ($4 \times 4 \times 1$) for all slabs supercells. DFT
calculations were performed for 11 transition metals from the groups 9th to
12th in the periodic table with either fcc (Rh, Ir, Ni, Pd, Pt, Cu, Ag, Au) or
hcp (Co, Zn, Cd) structure. The corresponding equilibrium bulk lattice
constants are shown in Tables S1 and S2 of the supporting information for fcc
and hcp metals, respectively.

\begin{figure}[!]
  \centering
  \includegraphics[width=\textwidth]{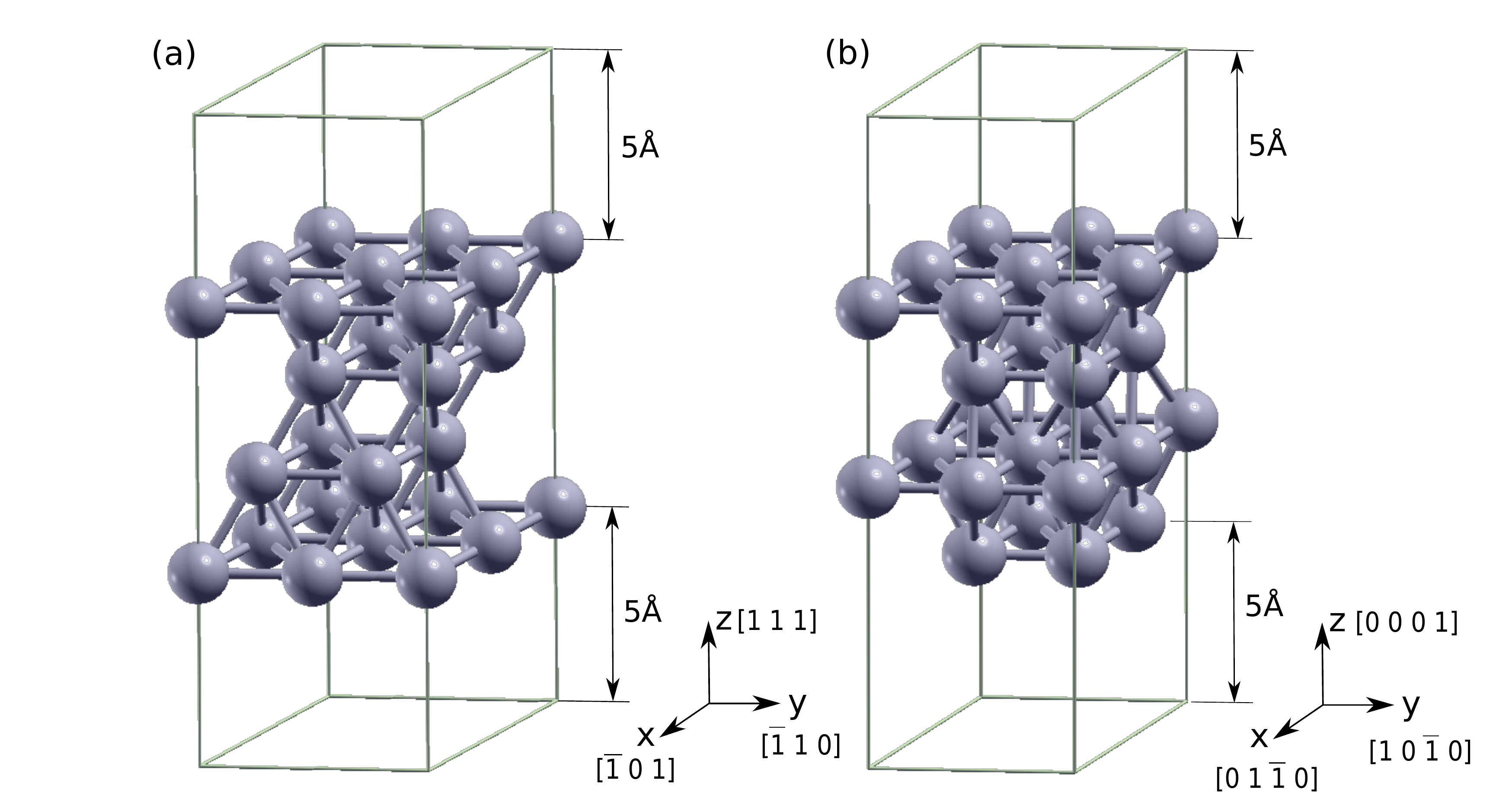}
\caption{(a) Four-layer slab supercell of a (111) fcc surface. (b) {\it Idem}
for a four-layer slab supercell of a  (0001) hcp surface. The surfaces in the
supercells are separated by 10\AA ~of vacuum.}
\label{supercells}
\end{figure}

The adsorption of H, O, and OH was modeled on four-layer slabs with the (111)
surface facet in the fcc metals and the (0001) surface facet in the hcp metals.  
 These surfaces were selected because they present the lowest surface energy -- and, thus, stand for the most stable ones -- for each lattice ~\cite{Jain2013}. The (2x2) slab supercells with four layers of atoms perpendicular to the
surface were generated with the Atomic Simulation Environment (ASE)
\cite{998641} from the equilibrium lattice parameters.  The periodic slabs were separated by 10 \AA~  of vacuum (Figure \ref{supercells}) in the direction perpendicular to the surface. Adsorption energies were
calculated of each chemical specie assuming a coverage of 1/4 monolayers.  All metal atoms in the top two layers of the slab and all adsorbed H, O, and OH
were fully relaxed, while the positions of metal atoms in the bottom two layers
were fixed. The adsorption energies of H in the case of Pt (111) surfaces were
evaluated for slabs including 3, 4, and 5 atomic layers perpendicular to the
surface and the results obtained with 4 and 5 layers were very close,
indicating that 4 layers were enough to reach convergence.  Moreover,
simulations of H adsorption on Pt (111) surfaces carried with ultrasoft
pseudopotentials were compared with those obtained using the projector
augmented-wave method (PAW), which provides more accurate results
~\cite{Blchl1994,Prandini2018,Fearon2006,Hanh2014,Kolb2014} with
higher computational cost.  The differences in adsorption energies between both simulations were
negligible. Thus, it was concluded that the combination of the PBE functional
with an ultrasoft pseudopotential in four-layer surface slabs offered the best
balance between accuracy and computational time. 

The adsorption energy of atomic H and O was calculated as

\begin{align}
E_\mathrm{adsX} = E_\mathrm{slab+X} - ( E_\mathrm{slab} + \frac{1}{2} E_\mathrm{X_{2}} )
\end{align}

\noindent where $E_\mathrm{slab}$ and $E_\mathrm{slab+X}$ stand for the total
energies of the slab without and with the absorbate X which represents H and O,
respectively. $E_\mathrm{X_{2}}$ accounts for the total energy of the hydrogen
and oxygen molecule in the gaseous state.  It should be noted that  the adsorption of O$_2$ from a molecule of H$_2$O (instead from O$_2$) should be used to calculate the catalytic activity but the difference from both adsorption energies is given by constant  that depends only on the formation energies of O$_2$, H$_2$ and H$_2$O and is  independent of the material of the slab and of the applied strain. Thus, isolated molecules of H$_2$ and O$_2$ were assumed to calculate the adsorption energies in this investigation.

The adsorption energy of OH was calculated as

\begin{align}
E_\mathrm{adsOH} = E_\mathrm{slab+OH} - ( E_\mathrm{H_{2}O} - \frac{1}{2} E_\mathrm{H_{2}} )-  E_\mathrm{slab}
\end{align}

\noindent where $E_\mathrm{slab+OH}$ stands for the total energy of the slab
with OH, and $E_\mathrm{H_{2}}$ and $E_\mathrm{H_{2}O}$ account for the total
energies of the hydrogen and water molecules in the gaseous state,
respectively.

The metal surfaces in Figure \ref{supercells} were subjected to normal and shear
stresses in the surface plane.  Mixed boundary conditions are imposed to solve the elastic problem in the DFT calculations. They include imposed strains in the slab plane and zero stresses on the free surface. The deformation gradient ${\mathbf F}$ applied
to the supercell was

\begin{equation}
\begin{array}{ccc}
{\mathbf F}= \begin{pmatrix}
1+\epsilon_1 & \gamma & 0\\
0 & 1+\epsilon_2 & 0\\
0 & 0 & 1
\end{pmatrix}
\end{array}
\end{equation}

\noindent where  $\epsilon_1$ and $\epsilon_2$ stand for the normal strains along $x$ and $y$ directions while and $\gamma$  for the shear distortion in the $xy$ plane. Uniaxial deformation was applied when $\epsilon = \epsilon_1$ and $\epsilon_2 = \gamma$ = 0, while $\epsilon = \epsilon_1 = \epsilon_2$ and $\gamma =0$ for biaxial deformation. 

In addition, the mechanical stability of surface slab supercells under strain
was analyzed from the harmonic phonon spectrum using the Phonopy code
\cite{phonopy}. To calculate the phonon spectrum, the atomic forces after
perturbing slightly the atomic positions from the equilibrium positions were
calculated for different strains in large slab supercells. They were obtained by repeating
the basic slab supercells in Figure \ref{supercells} by 2 times in the $x$
direction and two times in the $y$ direction, leading to supercells with 64
atoms. The number of perturbations to obtain the phonon spectrum depends on the
symmetries of the supercell which in turn depend on the applied strain  and can
vary from 4 (no strain applied) to 24 (15\% biaxial deformation).
It should be noted that the phonon calculations are computationally very
expensive and they were used to assess the maximum strain levels that should be
explored in the adsorption calculations under normal and shear strains because
mechanical instabilities are likely to appear well before these limits.

\section{Results} 

\subsection{Adsorption energies of H, O, and OH}

\begin{figure}[!t]
  \centering
\includegraphics[width=\textwidth]{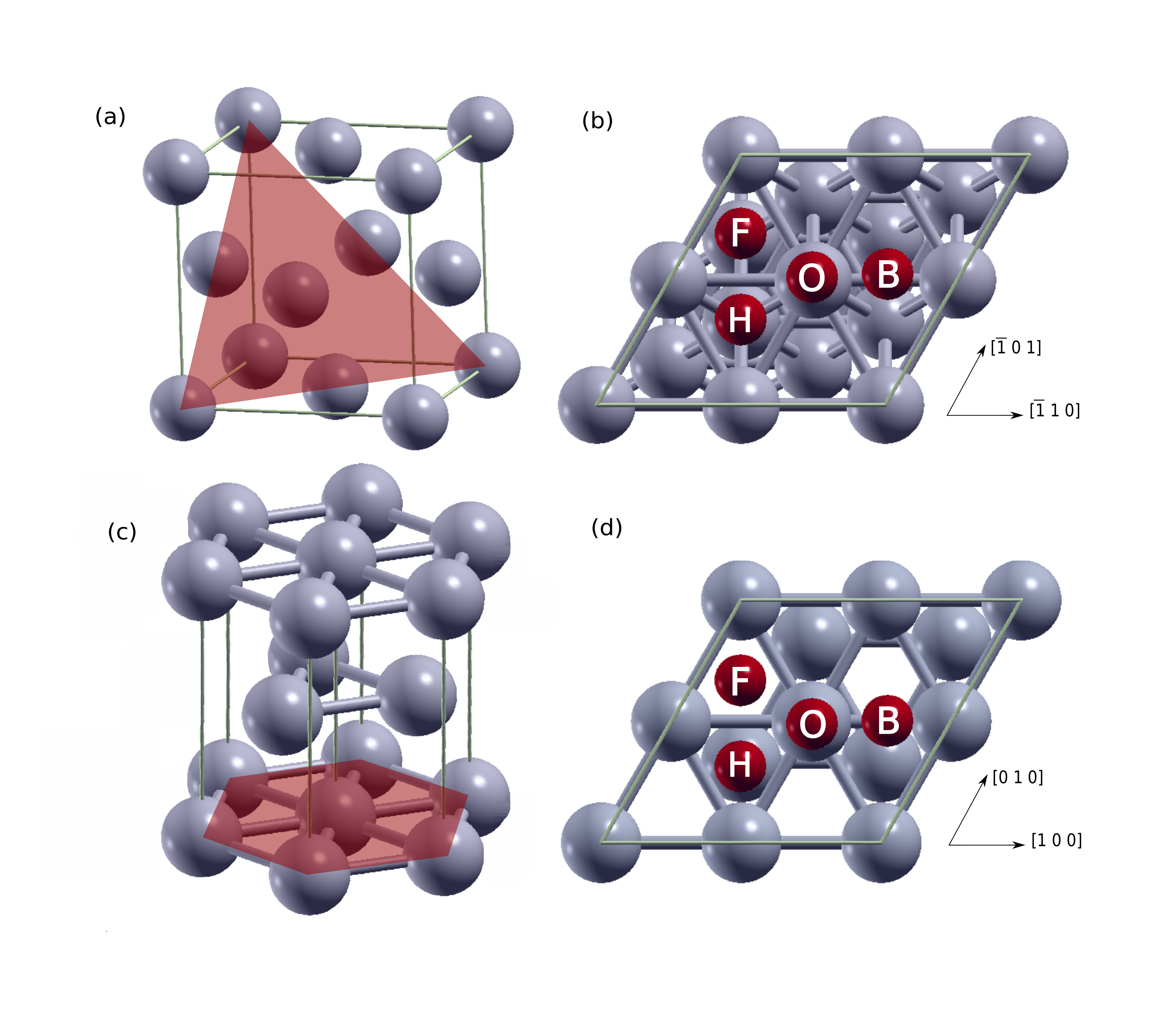}
  \caption{(a) fcc unit cell. The (111) plane is shadowed in red. (b) Perpendicular view to the (111) plane of the fcc lattice showing the binding sites
  for adsorbates. (c) hcp unit cell. The (0001) plane is shadowed in red. (d) Perpendicular view to the (0001) plane of the hcp lattice showing the binding
  sites for adsorbates. The red circles in (b) and (d) denote the binding
  sites and the labels 'H', 'F', 'O', and 'B' represent the HCP, FCC, ONTOP,
  and BRIDGE positions, respectively.}
  \label{junta}
\end{figure}

Adsorption energies were calculated in the absence of applied strains in all
the available positions for all studied metals. The four possible positions in
which H, O, and OH can be adsorbed onto the  (111) fcc and the (0001) hcp
 surfaces are shown in figure \ref{junta}(a-b) and (c-d), respectively. They are labeled H, F, O,
and B in the figure and stand for the HCP, FCC, ONTOP, and BRIGDE positions on
the surfaces, respectively. It was found that adsorbates at BRIDGE position
diffused towards more favorable positions and they were omitted from this
study.  In the case of OH adsorption, the molecule was placed perpendicularly
to the adsorption plane with the O atom closer to the surface.

The adsorption energies for H, O, and OH onto the surfaces are shown in Table
\ref{table:nostrain} for the different adsorption sites. There are very large
differences between the strongest (FCC) and the weakest (ONTOP) adsorption
sites in many cases. For instance, the adsorption of oxygen onto nickel is
associated with an energy of -3.13 eV in the FCC position but is reduced to
-1.20 eV on the ONTOP site. On the contrary, the differences in adsorption
energies between the FCC and HCP positions are much smaller. 
These trends are in agreement with previous experimental and theoretical results ~\cite{Nrskov2004,Nrskov2005,Pang2011,Lvvik1998}. For instance, the difference with the calculated adsorption energy of H/Co in ~\cite{Nrskov2005} was only 0.03 eV and 0.23 eV in the case of OH/Cu adsorption ~\cite{Nrskov2004}.  Moreover, The FCC position was found to be the most favorable adsorption
site (most negative energy) in the fcc metals while the FCC and HCP positions
had very similar adsorption energies in the hcp metals. The ONTOP site
presented the least favorable adsorption energy in all cases, except for
Ir.

Regarding the different adsorbates, the adsorption process of O is always
exothermic for the FCC and HCP positions whereas the adsorption of OH is an
endothermic process in most cases (with the exception of Co) with positive
adsorption energies. The adsorption energies for H are, in general, smaller in
absolute values and can be endothermic or exothermic depending on the
transition metal.

\begin{table}[h]
\centering
\caption{Adsorption energies of H, O, and OH on FCC, HCP, and ONTOP sites for
all the metals. The adsorption surface of the fcc metals (Ni, Cu, Pd, Ag, Pt,
Au, Rh, and Ir)  was the (111) plane, whereas adsorption occurred at the (0001)
plane in the hcp metals (Zn, Cd, and Co). The adsorption energies for OH in Zn
and Cd in the ONTOP position are not indicated because the molecule moves
towards the BRIDGE position. The energy values are expressed in eV.}
\small
\hspace*{-2cm}
\begin{tabular}{@{}lccc|lccc|lccc@{}}
\toprule
System    & FCC & HCP & ONTOP & System    & FCC & HCP & ONTOP & System & FCC & HCP & ONTOP \\ 
\midrule
H/Ni & -0.52   & -0.50   & 0.06      & O/Ni & -3.13   & -3.03   & -1.29  & OH/Ni & 0.03   & 0.57   & 1.04    \\
H/Cu & -0.25   & -0.23   & 0.39      & O/Cu & -2.68   & -2.54   & -0.82  & OH/Cu & 0.14   & 0.18   & 1.05    \\
H/Pd & -0.54   & -0.51   & -0.01     & O/Pd & -2.16   & -1.99   & -0.58  & OH/Pd & 0.71   & 0.85   & 1.64    \\
H/Ag & 0.17    & 0.18    & 0.73      & O/Ag & -1.38   & -1.27   &  0.13  & OH/Ag & 0.72   & 0.75   & 1.57    \\
H/Pt & -0.49   & -0.43   & -0.42     & O/Pt & -2.16   & -1.75   & -0.73  & OH/Pt & 1.19   & 1.50   & 1.91    \\
H/Au & 0.09    & 0.14    & 0.39      & O/Au & -0.98   & -0.72   &  0.50  & OH/Au & 1.52   & 1.58   & 2.22    \\
H/Rh & -0.53 & -0.53 & -0.25   & O/Rh & -2.96 & -2.84 & -1.55 & OH/Rh & 0.27 & 0.45 & 1.15  \\
H/Ir & -0.39 & -0.36 & -0.47 & O/Ir & -2.62 & -2.40 & -1.52 & OH/Ir & 0.76 & 0.98 & 1.29  \\
\midrule
H/Co & -0.54 & -0.51 & 0.11    & O/Co & -3.34 & -3.42 & -1.97 & OH/Co & -0.13 & -0.2 & 0.55 \\
H/Zn & 0.70  & 0.67  & 0.81    & O/Zn & -2.53 & -2.62 & -2.59 & OH/Zn & 0.46 & 0.37 & --- \\
H/Cd & 0.80  & 0.81  & 0.79    & O/Cd & -2.19 & -2.24 & -2.22 & OH/Cd & 0.36 & 0.31 & --- \\
\bottomrule
\end{tabular}
\label{table:nostrain}
\end{table}

\subsection{Effect of elastic strain on the adsorption energies of H, O, and OH}

\subsubsection{Stability limits}  

The mechanical stability limits of fcc Cu and Pt slabs under biaxial tension
and compression as well as pure shear were determined from the harmonic phonon
spectra. Calculations were limited to 5\%, 10\%, and 15\% biaxial strains in
tension, -5\% and -10\% biaxial strains in compression, and 10\% strain in shear
because of the high computational cost. The phonon density of states spectra
for the Pt and Cu slabs can be found in the supporting information (Figures S1
and S2). Negative frequencies appeared in the fcc slabs subjected to 15\%
biaxial tension in Pt and to 10\% biaxial tension in Cu. Similarly, negative
frequencies appeared in both metal surfaces at -10\% biaxial compression or
shear. In the case of  the (0001) slabs of the hcp metals, it was found
that that the order in the supercell was lost for small strains, indicating that the mechanical instabilities triggered by compressive or shear strains in (0001) hcp slabs develop  sooner than in (111) fcc slabs.

Taking into account the theoretical values of the stability limits obtained
with the phonon calculations, it was decided to explore the effect of elastic
strains on the adsorption energies in the (111) fcc slabs up to 8\% tensile
strain, -5\% compressive strain,  and 4\% shear strain. In the case of (0001)
hcp slabs, only the effect of tensile strains (and of small compressive strains in particular
cases) on the adsorption energy was analyzed.

\subsubsection{Influence of mechanical strains on the optimum adsorption site}

\begin{figure}[h!]
\hspace*{-1.5cm}
  \centering
  \includegraphics[width=1.2\textwidth]{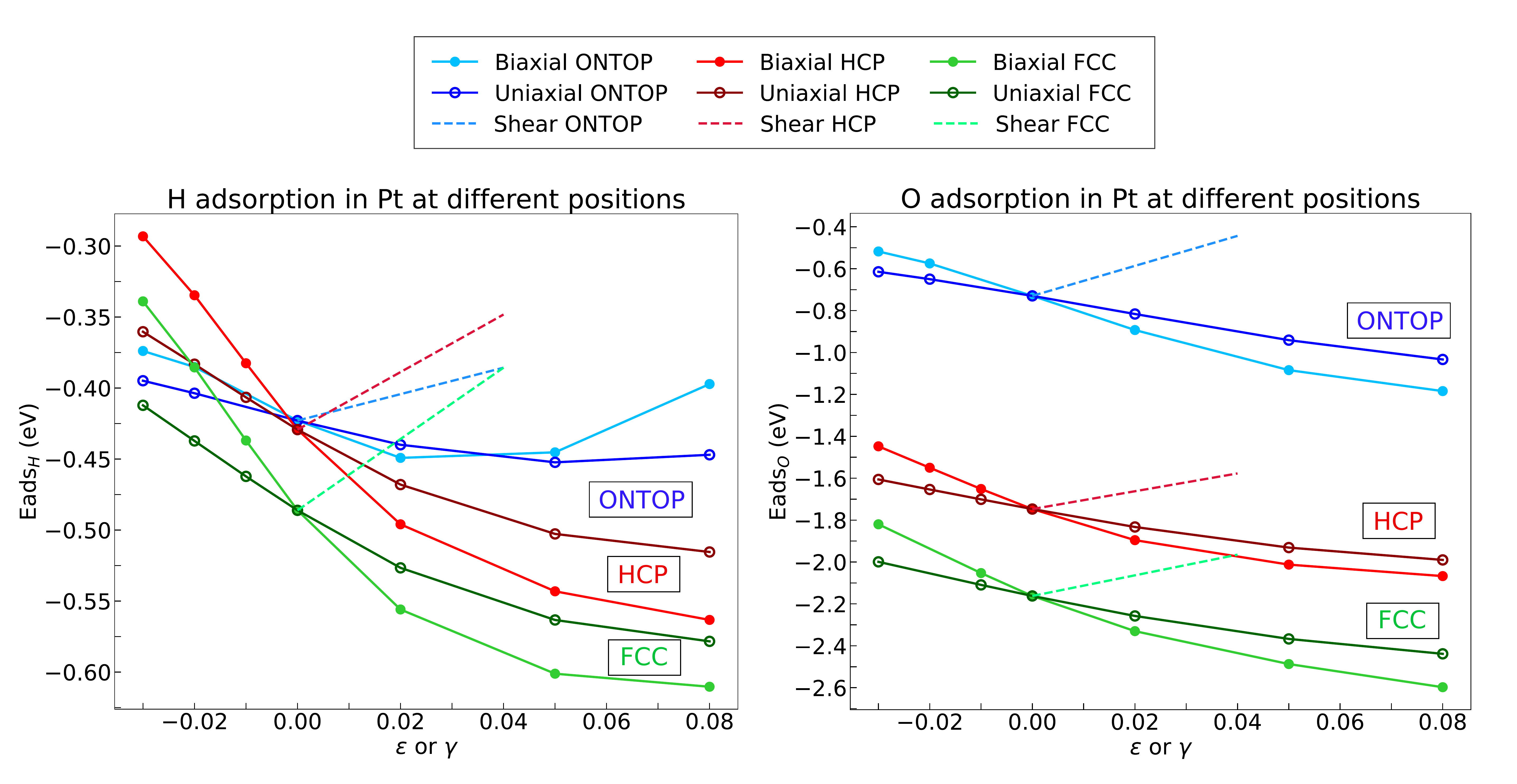}
  \caption{Effect of strains (either uniaxial, biaxial or shear) on the adsorption 
  energy of (a) H, $E\mathrm{ads_H}$ and (b) O,  $E\mathrm{ads_O}$ onto different sites of (111) Pt surfaces.}
  \label{Ptpositions}
\end{figure}

The effect of mechanical strains (either uniaxial, biaxial or shear) on the
adsorption energies of H and O onto the different adsorption sites (FCC, HCP,
ONTOP) were determined by DFT calculations in Pt. The results are plotted in
Figure \ref{Ptpositions} for either uniaxial  ($\epsilon = \epsilon_1$,
$\epsilon_2 = \gamma$ = 0) or biaxial   ($\epsilon = \epsilon_1=\epsilon_2$,
$\gamma$ = 0) deformations in the range -3\% up to 8\%, and for shear strains
$\gamma$ in the range 0\% to 4\% ($ \epsilon_1=\epsilon_2$ =0), in agreement
with the  limits indicated above. 

The same behavior can be observed for all the adsorption sites and adsorbates:
compressive strains increase the adsorption energy (less negative and,
therefore, adsorption is less favorable) while tensile strains lead to the
opposite behavior. Moreover, the variation in adsorption energy with strain is
always higher in the case of biaxial deformation. In addition, shear
deformations behave as compressive deformations and increase the adsorption
energy. The only exception to these trends is found in the adsorption energy of
H onto the ONTOP position subjected to very large biaxial tensile strains ($>$
5\%), which lead to a slight increase in the adsorption energy. This difference
may be caused by the proximity to the instability limits of the slab at
this strain. 

The effect of mechanical strains on the adsorption energy of O is higher than
that of H but it should be noted that the absolute values of the adsorption
energies are also much higher in the former. In addition, the effect of
mechanical strains is similar for all the adsorption sites for a given
adsorbate. Thus, application of mechanical strains does not change the most
favorable adsorption site for H and O in Pt, which is always FCC. Similar
results were obtained for the (111) surfaces of other fcc metals, while the
optimum adsorption site for (0001) surfaces of hcp metals is the HCP and it
is also independent of the strain state.

\subsubsection{Influence of mechanical strains on the adsorption energy of H, O and OH}

The analysis of the influence of mechanical strains on the adsorption energy of
H, O, and OH was focussed in the FCC sites of (111) fcc surfaces and in the HCP
sites of the (0001) hcp surfaces, which are the most favorable locations for
adsorption.

\begin{figure}[t!]
\hspace*{-1.5cm}
  \centering
  \includegraphics[width=1.2\textwidth]{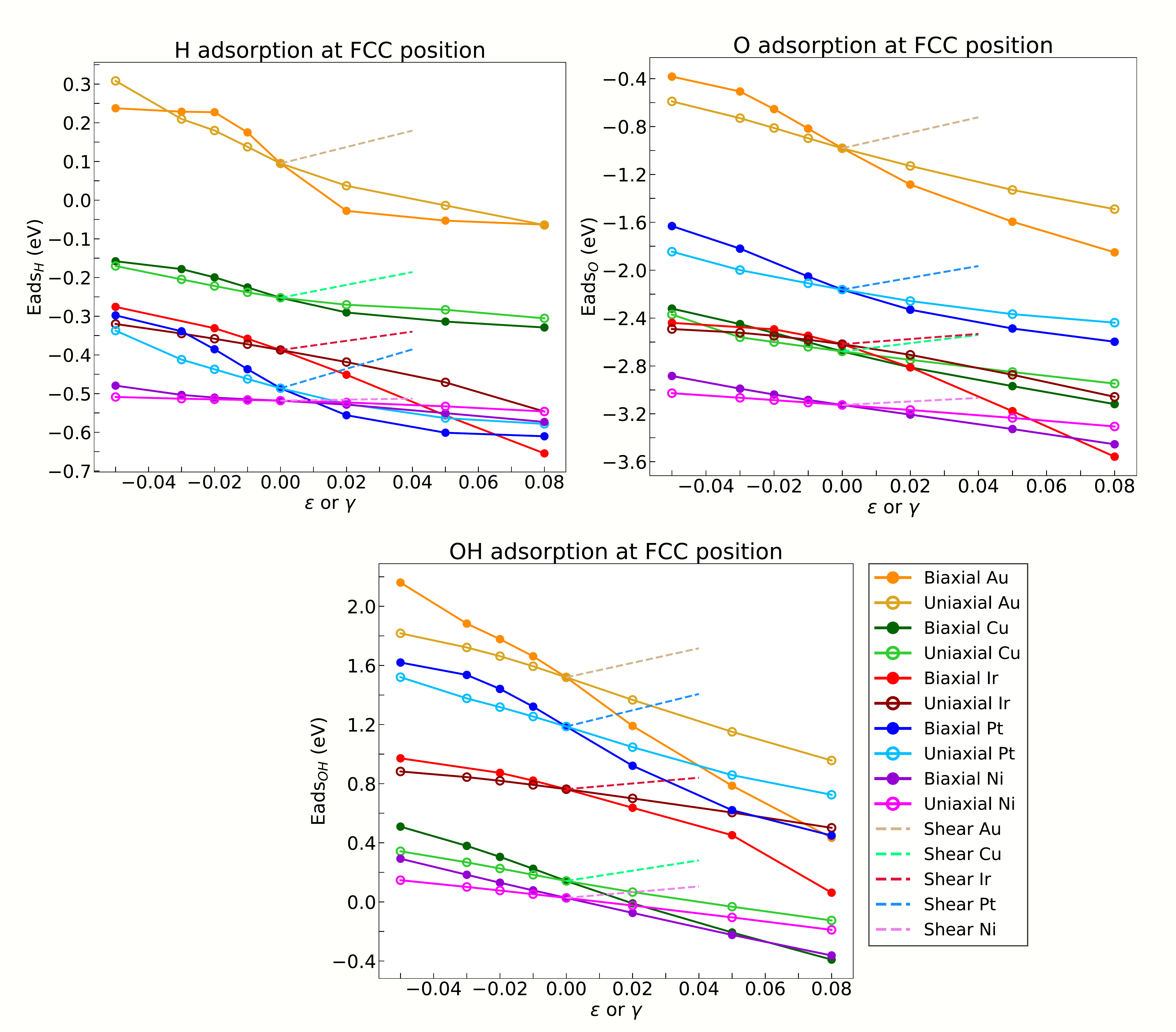}
  \caption{Effect of strain (either uniaxial, biaxial or shear) on the
  adsorption energy of (a) H, $E\mathrm{ads_{H}}$; (b) O,  $E\mathrm{ads_{O}´}$
  and (c) OH, $E\mathrm{ads_{OH}´}$ onto the FCC sites of (111) surfaces of fcc
  transition metals. Curves for Ag, Pd, and Rh are omitted because they overlap
  with the curves for Au, Pt, and Ni, respectively. Separate figures of all
  metals can be found in Figures S3 to S8 of the supporting information.}
  \label{Figure_1y2}
\end{figure}

\begin{figure}[t!]
\hspace*{-1.5cm}
  \centering
  \includegraphics[width=1.2\textwidth]{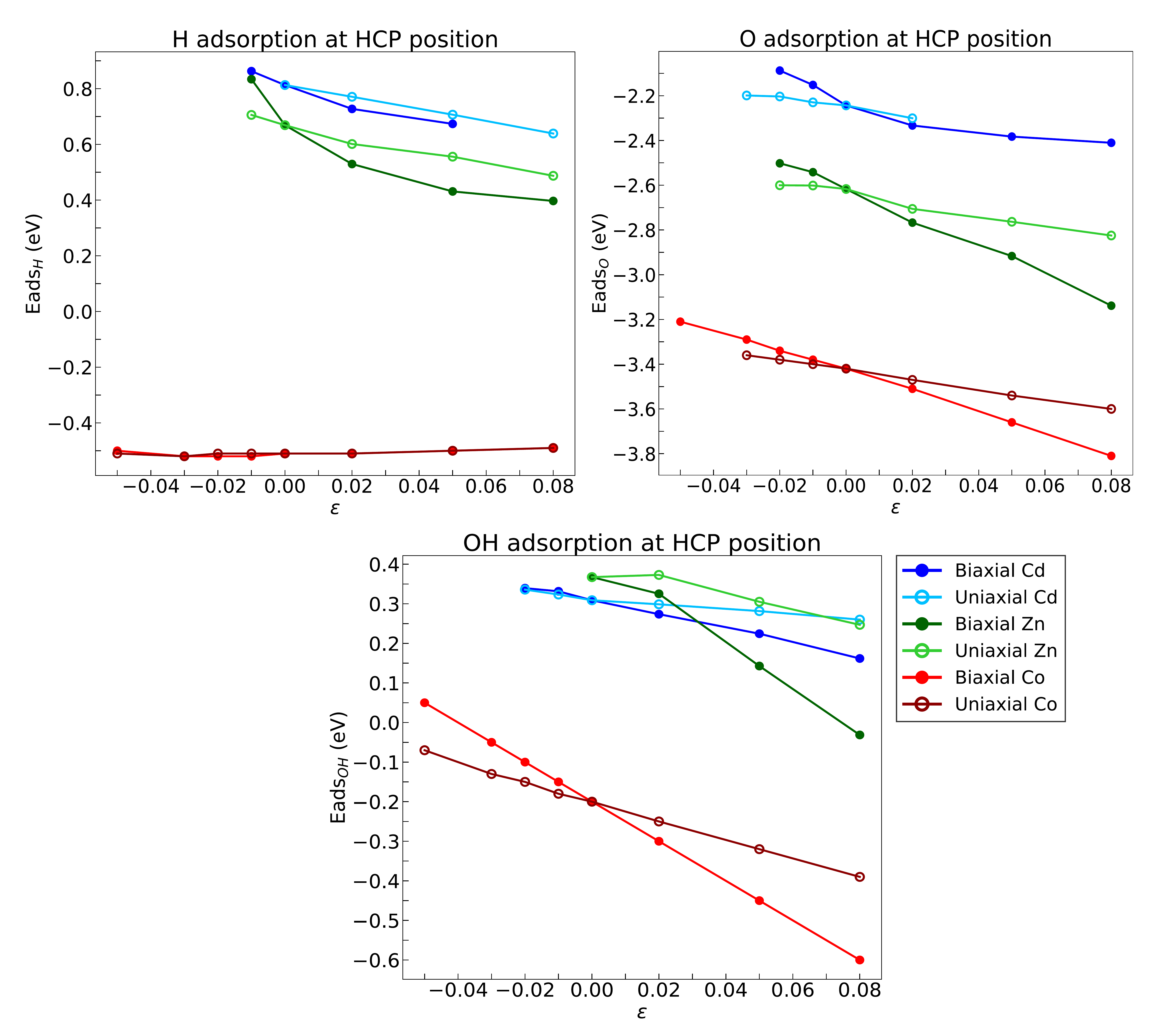}
  \caption{Effect of strains (either uniaxial or biaxial) on the adsorption
  energy of (a) H, $E\mathrm{ads_H}$; (b) O,  $E\mathrm{ads_O}$  and (c) OH,
  $E\mathrm{ads_{OH}}$ onto the HCP sites of (0001) surfaces of hcp transition
  metals.}
  \label{nueva}
\end{figure}

The effect of mechanical strain (either uniaxial, biaxial or shear) on the
adsorption energies of H, O, and OH on the FCC sites of the (111) surfaces of
eight fcc transition metals (Cu, Pt, Ni, Au, Ir, Rh, Pd, and Ag) is plotted
in Figure \ref{Figure_1y2}. The corresponding results on the HCP sites of the
(0001) surfaces are plotted for Co, Zn, and Cd in Figure \ref{nueva}. As
indicated above, application of compressive and shear strains was restricted in
hcp metals because of the development of instabilities in the supercell.  

The highest (less favorable) adsorption energies are always found for OH while
the lowest (more favorable) are reported for O. Adsorption energies for H are
smaller (in absolute values) than those calculated for O and OH  in all metals,
following the trends reported above for Pt. In addition, the adsorption
energies increase (become less negative) with compressive strains and decrease (become more
negative) with tensile strains, while biaxial deformations have stronger
influence than uniaxial deformation on the adsorption energy for the same
strain. Shear strains increase slightly the adsorption energies in all cases.
For a given metal, the effect of mechanical strains on the adsorption energy of
O and OH quite similar, very likely because adsorption is dominated by the
larger O atom. 

The variation in adsorption of energy of O and OH with strain is significantly
higher than that of H and these differences can be attributed to the larger
atomic radius of the O atom. The additional p-orbitals increase the size of the
electronic environment and, therefore, the overlap with the d-band structure of
the metals. The presence of p-electrons further apart from the nucleus favors
the interaction with the d-electrons of the surfaces.  

Indeed, the effect of mechanical strains on the adsorption energies depends on
the metal. Noble metals, such as Au and Pt, show the highest sensitivity to the
strain while Ni shows much lower sensitivity. It is also worth noting that the
adsorption of H onto Co is not affected by either uniaxial or biaxial strains
in the range - 5\% to 8\% but the largest changes in the adsorption energy of O
and OH  ($>$ 0.5 eV) are found in Co for the same strain range.

The trends observed for the adsorption energies in this work are in agreement
with the predictions of the d-band theory. This theory states that the more
favorable adsorption energies related with tensile and the less favorable ones
corresponding to compressive and shear strains can be explained in terms of
displacements in the d-band center \cite{Hammer2000, Hammer1995, Pang2011,Mavrikakis1998}. Indeed, the d-band theory indicates that  tensile strains shift the d-band center towards higher energies for transition metals with more than half-filled d-bands. A higher d-band center results in stronger bonding and leads to more favorable adsorption energies, while compression and shear strains produce a shift towards a lower d-band center and lead to less favorable interactions.

\subsection{Relationship between adsorption energy and hole area}

\begin{figure}[t!]
\hspace*{-1.5cm}  \centering
  \includegraphics[width=1.2\textwidth]{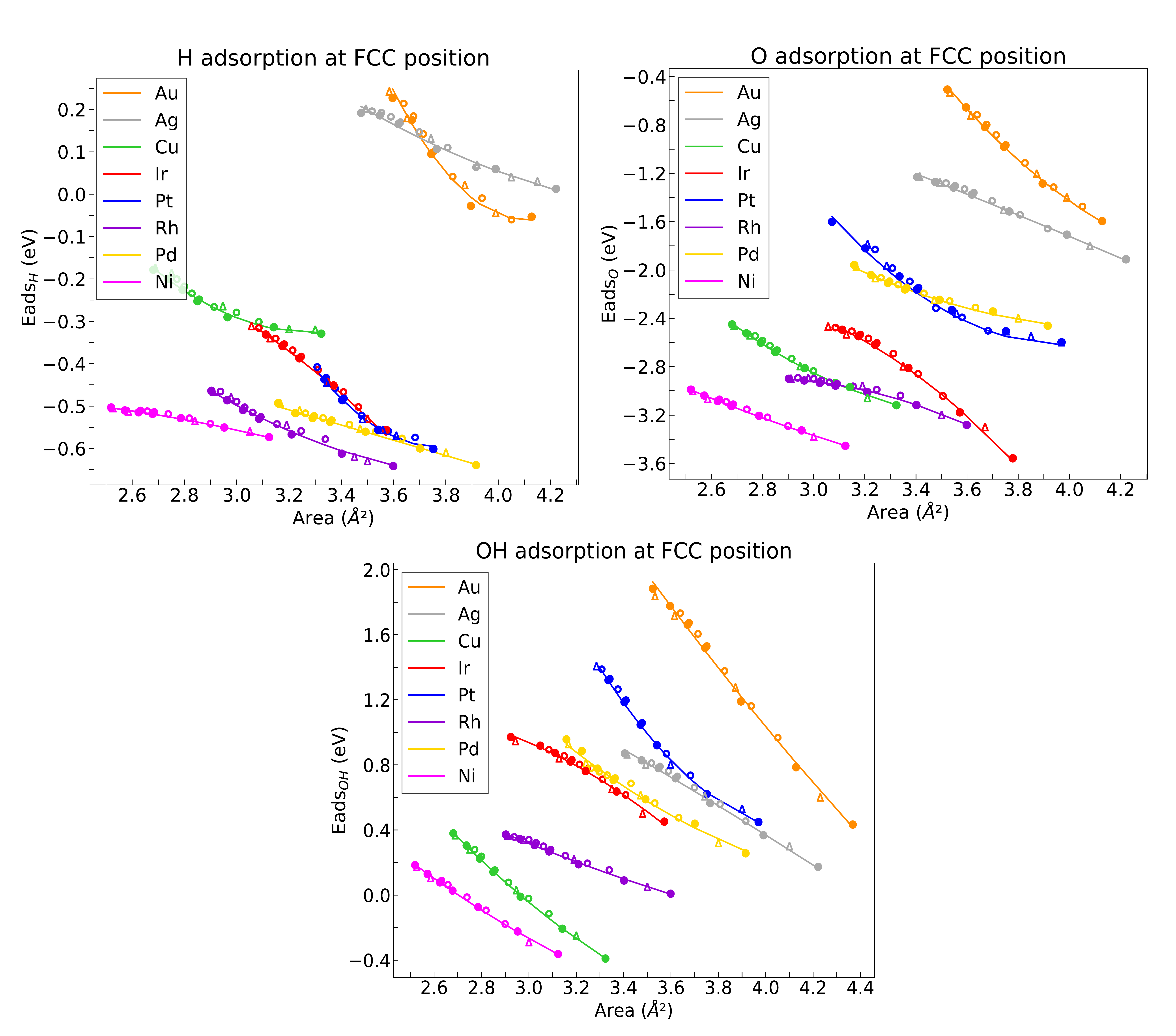}
  \caption{Adsorption energies as a function of the area of the adsorption hole
  that is function of the applied deformation gradient. (a) H adsorption at FCC
  positions of (111) surfaces of fcc transition metals. (b) O adsorption at FCC
  positions of (111) surfaces of fcc transition metals. (c) OH adsorption at
  FCC positions of (111) surfaces of fcc transition metals. Solid and open
  circles stand for the adsorption energies calculated under biaxial and
  uniaxial strains, respectively, while open triangles refer to adsorption
  energies under shear strains or a combination of shear and axial strains. The
  lines show the best fit to the DFT results in the strain ranges indicated in
  Figure \ref{Figure_1y2}.}
  \centering
  \label{AREAS1}
\end{figure}

\begin{figure}[t!]
\hspace*{-1.5cm}
  \centering
  \includegraphics[width=1.2\textwidth]{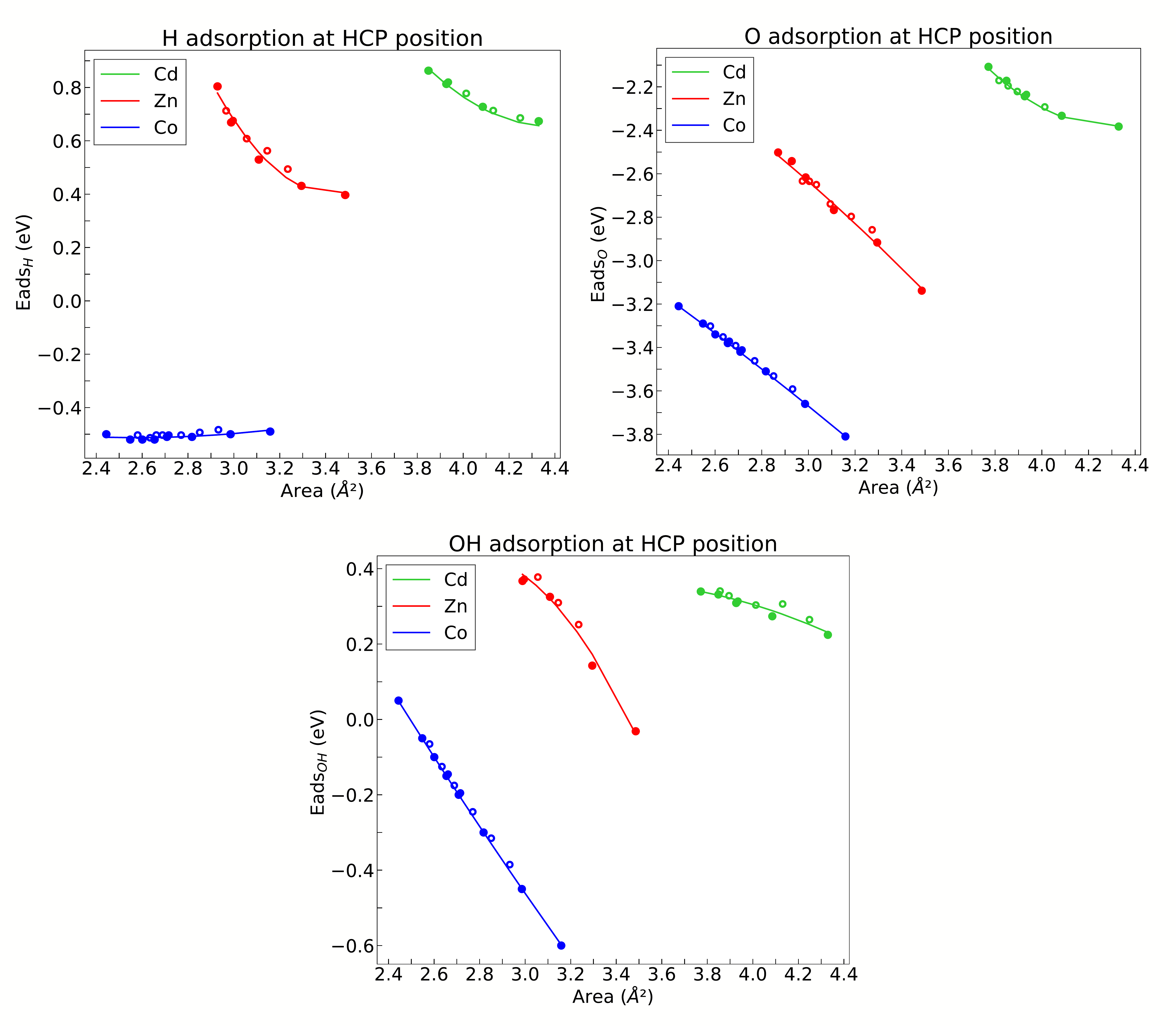}
  \caption{Adsorption energies as a function of the area of the adsorption hole
  that is function of the applied deformation gradient.  (a) H adsorption
  at HCP positions at (0001) surfaces of hcp transition metals. (b) O adsorption
  at HCP positions at (0001) surfaces of hcp transition metals. (c) OH
  adsorption at HCP positions at (0001) surfaces of hcp transition metals.
  Solid and open circles stand for the adsorption energies calculated under
  biaxial and uniaxial strains, respectively, and lines show the best fit to the DFT results in the strain ranges
  indicated in Figure \ref{nueva}.}
  \centering
  \label{AREAS2}
\end{figure}

While the d-band model provides a qualitative explanation of the trends
reported in Figures \ref{Figure_1y2} and \ref{nueva}, models that are able to
quantify the effect of mechanical strains on the adsorption energy are lacking.
In this respect, the adsorption energies of H, O, and OH in each fcc transitions
metals and in three hcp transitions metals are plotted in Figures \ref{AREAS1}
and \ref{AREAS2}, respectively, as a function of the area of hole (either FCC
or HCP) at which adsorption takes place. The area of the hole was calculated
from the length of the sides of the triangle which conform the FCC and the HCP
adsorption sites (Figure \ref{junta}(b) and (d)), which depends on the
deformation gradient $\bf{F}$ applied to the supercell. After relaxation, the
area of the hole was measured again using the code XCrysDen\citep{Kokalj1999}
and it was found that the differences in the hole area between the unrelaxed
and relaxed structure are negligible. 

The adsorption energies under biaxial strains (solid circles), uniaxial strains
(open circles) and a combination of shear and axial strains (open triangles)
are plotted for each metal in these figures. They show that the actual
adsorption energy only depends on the area of the hole and that the effect of
different strains can be superposed, i.e. different combinations of normal and
shear strains that lead to the same in hole area have the same effect of the
adsorption energy. Moreover, the relationship between the adsorption energy and
the hole area is fairly linear in most cases (although the DFT results are
better approximated by a parabola in several cases). Finally, the data in
Figures \ref{AREAS1} and \ref{AREAS2} allow to make a fast and accurate
estimation of the effect of mechanical strains on the adsorption energies of H,
O, and OH for any of these transition metals from the geometrical analysis of
the change in either FCC or HCP hole area under the application of a
deformation gradient $\bf{F}$. Because the hole area does not change during
relaxation, it is not necessary to perform the DFT simulations to determine the
adsorption energy.

The curves corresponding to the 1st row transition metals (Cu, Ni, Zn, and Co)
in Figures \ref{AREAS1} and \ref{AREAS2}  are located in the region with
smaller hole areas  because of the smaller size of these atoms. In addition, Ni
and Co also show very low adsorption energies, which can be attributed to
magnetism, while the adsorption energies of Cu and Zn are much higher for
similar hole areas. The curves corresponding to the 2nd and 3rd row transition
metals (Au, Ag, Ir, Pt, Rh, Pd, and Cd) are spread towards the right in Figures
\ref{AREAS1} and \ref{AREAS2} as a result of the larger atom size. In general,
the adsorption energies becomes more positive (less favorable)  when going from
left to right in the periodic table periods. This tendency is related with the
electronic density of the metals, whose d-bands become more populated as atomic
number of the metal increases in a period. Thus, as a general rule,  the
adsorption process  is favored by a reduction in the number of valence
electrons in the metal.

In most cases, the relationship between the adsorption energy and the hole area
is linear, although a parabola is more appropriate in several
cases. Moreover, the slope of the linear fit is similar for most transition
metals and large differences are only found in the case of H adsorption on Ni,
which is almost insensitive to the applied strains. These results indicate
that the mechanisms  that control the adsorption of H, O, and OH are very
similar and  the differences that appear in these figures  are ultimately
related to the particularities of d-orbitals of each metal and to magnetic
effects.

Finally, the adsorption energies of H and O onto the HCP position of (111) Pt
surfaces were determined for different magnitudes of axial strains to check
whether the relationship between adsorption energy and hole area could be
extrapolated to other adsorption sites. They are plotted in Figure S9 in the
supporting information, together with the adsorption energies onto the FCC
positions of (111) Pt surfaces, support the previous findings: for a given
adsorption site, the effect of mechanical strains on the adsorption energy of
transition metals only depends on the area of adsorption hole.

\section{Discussion} 

\subsection{Electronic structure}

\begin{figure}[t!]
\hspace*{-1.5cm}
  \centering
  \includegraphics[width=1.2\textwidth]{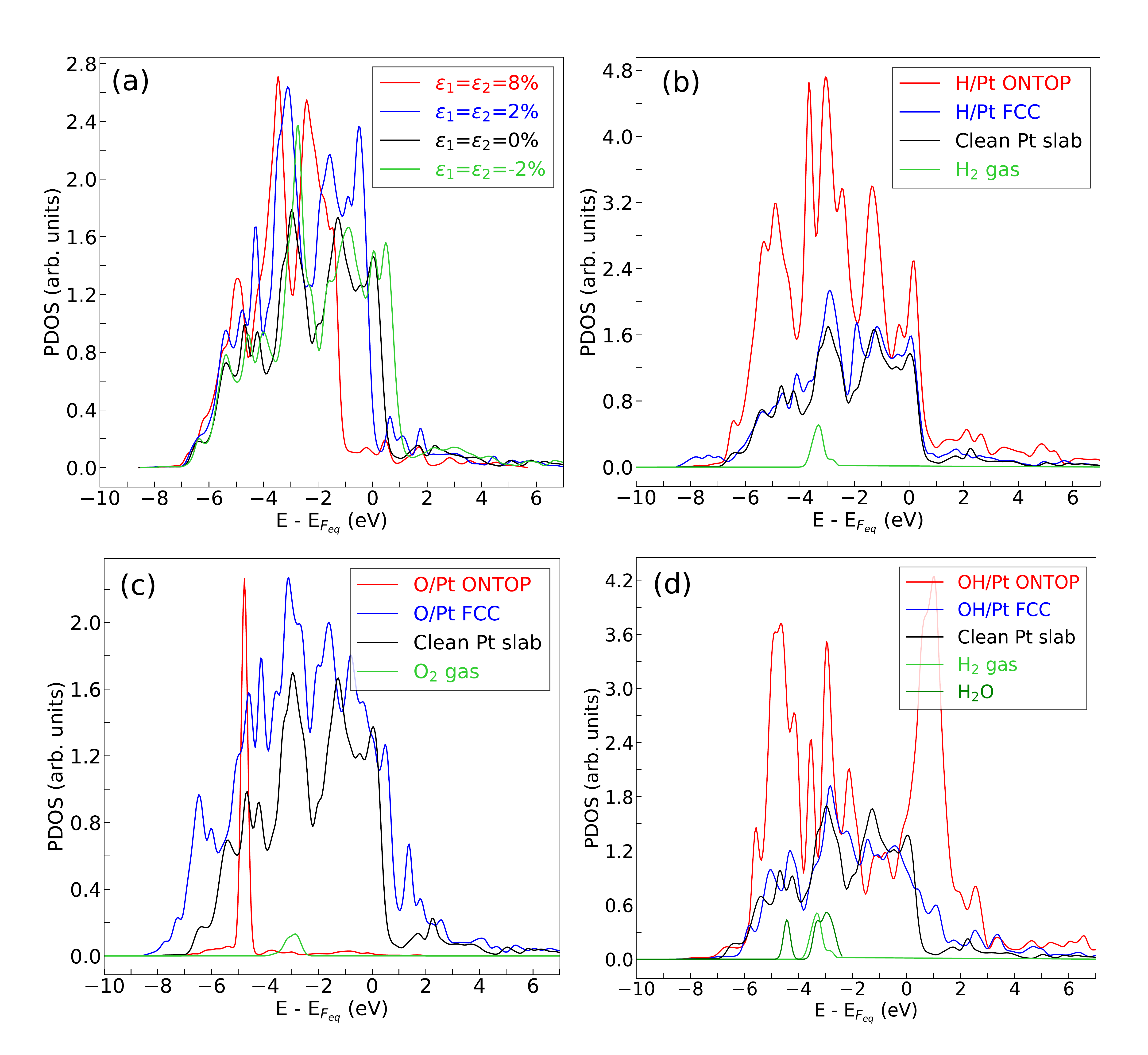}
  \caption{(a) PDOS onto the 5d orbitals of the (111) fcc Pt surface slab
  subjected to biaxial strain states characterized by $\epsilon$ = -2\%, 0\%,
  2\%, and 8\%. (b), (c), and (d) PDOS of the 5d orbitals of the (111) fcc Pt surface
  with a H, O, and OH, respectively, adsorbed at FCC and ONTOP positions. The
  corresponding PDOS of the  5d orbitals of the clean (111) fcc Pt slab, the
  H$_{2}$ molecule, the O$_{2}$ molecule, and the H$_{2}$O molecule at gaseous state are also included
  for comparison.  The energy values are referenced to the Fermi level of the
  (111) Pt slab when $\epsilon$ = 0 ($E_F$ = 3.5668 eV).}
  \label{Figure_6y7}
\end{figure}

It is known that changes in the surface geometry are accompanied by changes in the surface electronic structure \citep{Hammer2000}. To quantify these effects, the projected density of the states (PDOS) on the d-band onto the surface of all metals and adsorbates was analyzed to  explore the electronic origin of the
adsorption energy - hole area relationship. The PDOS corresponding to the (111) fcc Pt slab subjected to biaxial tensile  ($\epsilon$ = 2\% and 8\%)  or compressive  ($\epsilon$ = -2\%) strains  are plotted in Figure
\ref{Figure_6y7}(a)  and compared with the PDOS at $\epsilon = 0$.  The overlap of metal d-states at neighboring sites will either increase or decrease when a surface undergoes compressive or
tensile strains,  and so will the d-bandwidths in order to maintain a constant filling. Thus, compressive or tensile strains lead to downshifts or up-shifts of the d-band centers, respectively \cite{Kattel2014}, as shown in figure \ref{Figure_6y7}(a).

\begin{figure}[h!]
\hspace*{-1.5cm}
  \centering
  \includegraphics[width=1.2\textwidth]{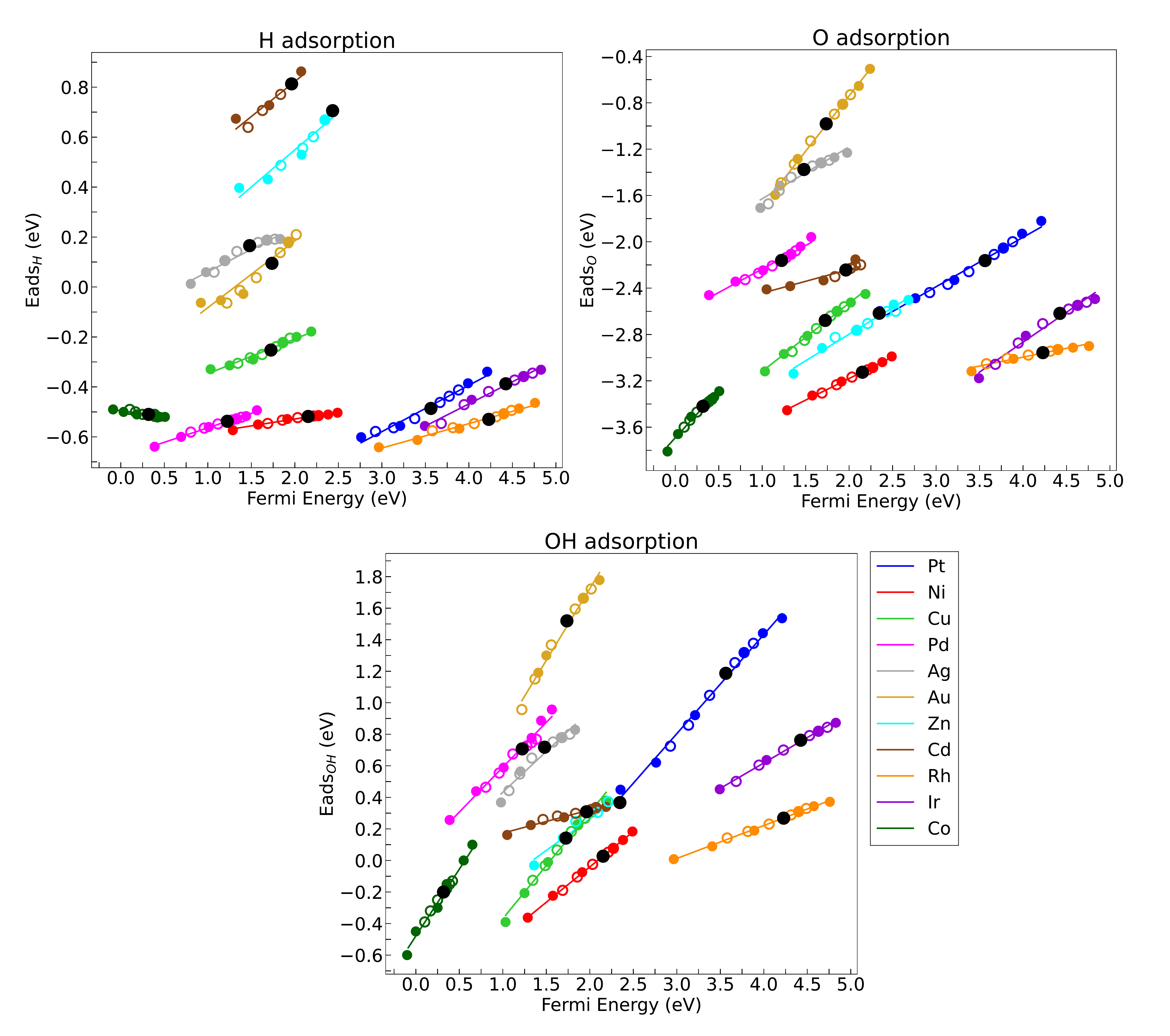}
  \caption{Adsorption energies of strained slabs as a function of the Fermi
  level in the clean, strained slab for all metals. (a) H adsorption. (b) O
  adsorption. (c) OH adsorption. The adsorption energies were calculated onto
  the FCC sites on (111) fcc slabs and onto HCP sites of (0001) hcp slabs.
  Values in both axes are expressed in eV. Solid and open circles represent
  the biaxial and uniaxial DFT calculations, respectively. The solid lines stand for the fit of the DFT results with eq. \eqref{AdsorptionFermi}. Black circles indicate the adsorption energy without strain.}
  \label{ADS_FERMI}
\end{figure}

The PDOS onto the 5d orbitals of a (111) fcc Pt surface after the adsorption of
a H atom onto the FCC site and the ONTOP site are plotted in Figure
\ref{Figure_6y7}(b), together with the PDOS of a hydrogen molecule  and of the
clean (111) fcc Pt slab. The adsorption of a H atom on the ONTOP position
generates strong electronic changes, which in turn are reflected in the PDOS.
In contrast, the PDOS when adsorption takes place onto the FCC site remains
practically identical to the one corresponding to the clean (111) Pt slab, with
the only difference being a small shoulder at $-$2 eV. This variation in the
PDOS can be considered negligible, as compared with the electronic changes
produced by the application of strain and, therefore, allows to establish a
direct link between the adsorption energy with the geometrical changes induced
at the FCC adsorption site by the application of strain.  The PDOS of a (111) fcc Pt surface after the
adsorption of an O atom onto the FCC site and the ONTOP site together with the
PDOS of an oxygen molecule, and of the clean (111) fcc Pt slab are plotted in Figure
\ref{Figure_6y7}(c) and very similar trends are observed.   The PDOS is localized in a single peak on the left after the adsorption of the O atom on the ONTOP position, whereas  the PDOS become broader but maintains its original shape after the adsorption of the O atom onto the FCC position.  Similarly, the PDOS of a (111) fcc Pt surface after the adsorption of OH onto the FCC site and the ONTOP site are plotted in Figure \ref{Figure_6y7}(d) together with the PDOS of a hydrogen and a water molecule, and of the clean (111)
fcc Pt slab. The trends are equivalent to those observed in Figs. \ref{Figure_6y7}(b) and (c). While the adsorption of OH at ONTOP position produces significant changes in the PDOS, the adsorption of OH at FCC position leads to negligible changes in the PDOS as compared with that of the clean (111) Pt slab. The above results suggest that the analysis the electronic structure of the
clean surface slabs may be enough to determine the effect of strain on the
adsorption energy of H, O, and OH in the transition metals studied in this investigation.
Moreover, the application of strain does not modify the shape of the PDOS
curves but only leads to a shift of the energy levels (Figure \ref{Figure_6y7}(a)). Thus,  the Fermi level  could be used as a metric of the electronic changes in the slabs upon the application of mechanical strains and, eventually, of the
adsorption energies. This hypothesis is supported by the results in  Figures
\ref{ADS_FERMI}(a), (b), and (c) in which the adsorption energies as a function
of the mechanical strains are plotted as a function of the Fermi level of the
clean, strained surface slabs for H, O, and OH adsorbates, respectively.  The
linear correlation between both magnitudes is obvious for all metals and
adsorbates and  can be represented by

\begin{equation}
E_\mathrm{adsX} = b +m E_F
\label{AdsorptionFermi}
\end{equation}

\noindent where X = H, O or OH and the coefficients $b$ and $m$ for each pair of adsorbate and transition metal can be found in Table S3 in the supporting information. It should be noted that $E_F$ in eq. \eqref{AdsorptionFermi} stands for the Fermi energy of the clean, strained slab. Obviously, these similarities indicate that the  underlying adsorption processes are governed by the same physical mechanisms.

\begin{figure}[h!]
  \centering
  \includegraphics[width=0.8\textwidth]{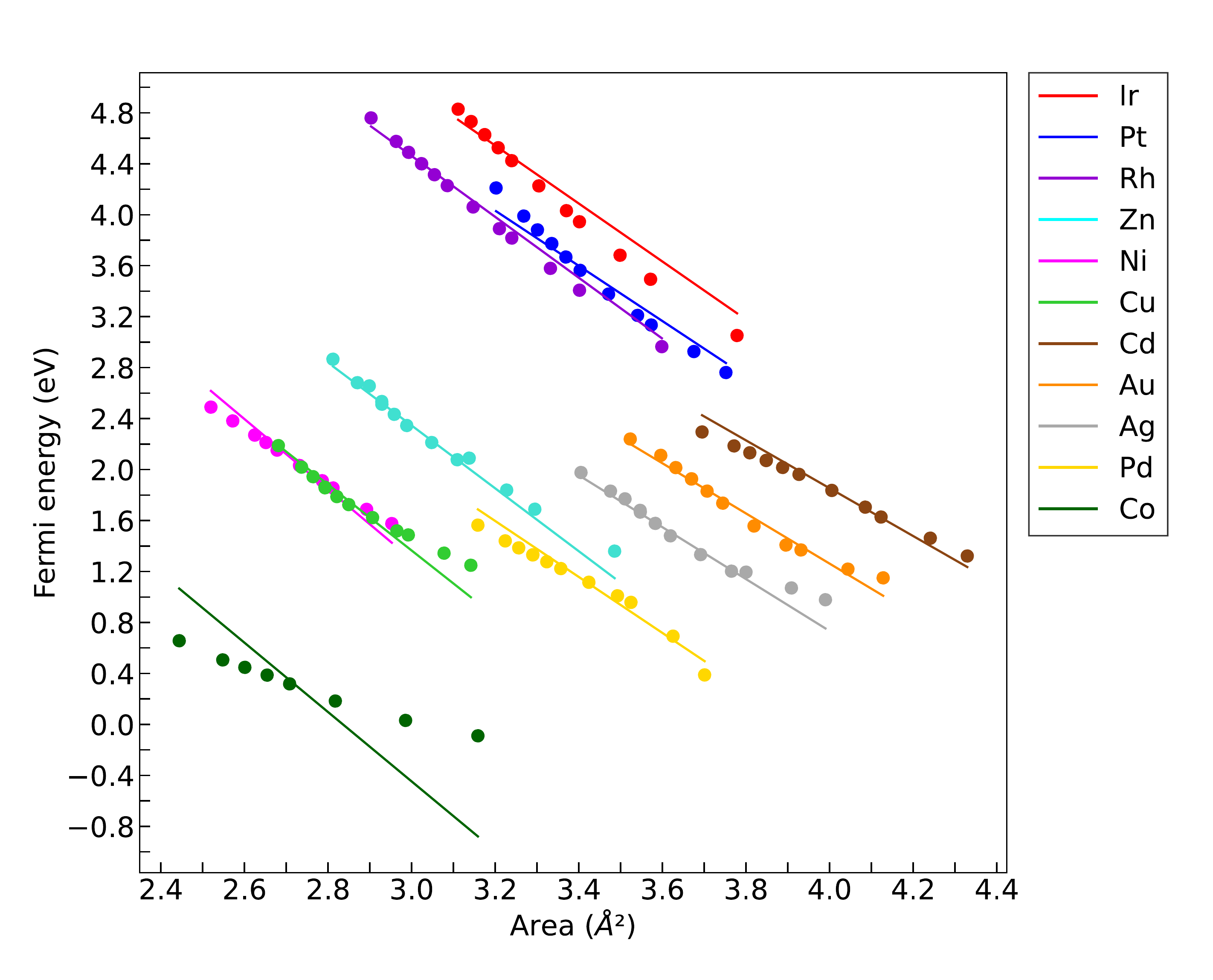}
  \caption{Evolution of the Fermi energy, $E_F$, at the surface of the
  different transition metals as a function of the area of the adsorption site,
  $A$. Circles stand for the results of the Fermi levels corresponding to the clean, deformed slabs  obtained  by DFT and the straight lines stand for the predictions of equation \eqref{FermiArea}.}
\label{fermi}
\end{figure}

The last step to link the adsorption energy with the area of adsorption site is
to find the relationship between the latter and the Fermi energy. The values of
both magnitudes obtained by DFT calculations on clean, strained slabs of the 11
transition metals are plotted in Figure \ref{fermi}. The straight lines in this
figure are given by

\begin{equation}
E_F = 7.4 -7.37 A + E_F(A_0)
\label{FermiArea}
\end{equation}

\noindent where $E_F$ is the Fermi energy (in eV), and $A$ the area of the
adsorption site in the surface subjected to a given strain state. $E_F(A_0)$ is
the Fermi energy of the surface corresponding to the undeformed state, which is given in Table S4 in the supporting information for each transition metal. This equation captures the independent contribution of the metal ligand (expressed by 
$E_F(A_0)$) and of the mechanical strain (given by $A$) to the Fermi energy and, thus, through eq. \eqref{ADS_FERMI} to the adsorption energy for each adsorbate. It should be noted the excellent agreement of this simple linear equation with the DFT calculations for most transition metals 
indicates that changes in the local electronic environment as a result of strain are better represented by the area of the hole than by the distance between one atom and the different neighbours.  Moreover, eq. \eqref{FermiArea} provides the explanation of the link between the adsorption energy and the area of the adsorption site in Figures \ref{AREAS1} and \ref{AREAS2}. The different behavior of Co -that does not follow equation \eqref{FermiArea}-  may be due to the magnetic properties of this metal but
further research is needed to clarify this point. Moreover, it should be noted
that preliminary studies on (110)  bcc Cr and (100) bcc V surfaces did not find
a clear relationship between adsorption energy, Fermi energies, and area of the
adsorption site. These differences may be attributed to the nature of the
d-orbitals in these metals, that are less than half or half filled in metals on
the left side of the periodic table. Thus, further research is needed to reach
a full understanding of the effect of mechanical strains on the adsorption
properties of transition metals.

\section*{Conclusions}

The influence of elastic strains on the adsorption of H, O, and OH on the (111)
surfaces of 8 fcc (Ni, Cu, Pd, Ag, Pt, Au, Rh, Ir) and on the (0001) surfaces
of 3 hcp (Co, Zn, Cd) transition metals was analyzed by means of DFT
calculations. The surface slabs were subjected to different strain states
(uniaxial, biaxial, shear, and a combination of them) up to strains dictated by
the mechanical stability limits indicated by the phonon calculations. It was
found that tensile strains favored the adsorption of the three adsorbates while
compressive and shear strains increased the adsorption energy (less negative)
and, thus, hindered the adsorption, in agreement with the d-band theory. Adsorption energies for H were smaller in absolute values than those calculated for O and OH  in all cases. Moreover, the optimum
adsorption (lowest energy)  of the three species was found onto the FCC sites
of the (111) fcc surfaces and onto the HCP sites of the (0001) hcp surfaces and
did not change with strain.

It was found that the variation of the adsorption energy in all metals due to
the application of mechanical strains  was only a function of the change in the
area of the adsorption site and the relationship between both magnitudes was
fairly linear in most cases. Thus, different combinations of normal and shear
strains that lead to the same change in the area of the adsorption site have
identical effect  on the adsorption energy. This general behavior indicated
that the physical mechanisms of adsorption were equivalent in all metals. The
analysis of the electronic structure showed that the application of strains did
not modify the  shape of PDOS of the d-orbitals of the transition metals but
only led to a shift in the energy levels. Moreover, the adsorption of H and O
on the surfaces led to negligible changes in the PDOS. Thus, the adsorption
energies of all adsorbates in all metals were a function of the Fermi energy
which in turn was associated to the change of the area of the adsorption
through linear law that was valid for all metals. 

As the change in the area of the adsorption site due to the application of
strain can be accurately determined by purely geometrical considerations, the
information in this paper allows the immediate and accurate estimation of the
effect of any elastic strain on the adsorption energies of H, O, and OH on 11
transition metals with more than half-filled d-orbitals. 
 This information can be used to predict the activation free energies of the different intermediates in the HER and ORR (as well as in other catalytic reactions) as a function of the applied strain for different transition metals with very limited computational cost, indicating the optimum combination of material and strain to enhance the catalytic activity. Moreover, the results in this paper can spur the search for correlations between geometrical descriptors of the elastic deformation and adsorption energies that can be used to make accurate predictions of the latter with minimum computational cost for other compounds and adsorbates.

\section*{Acknowledgments}

This investigation was supported by the MAT4.0-CM project funded by the Madrid
region under program S2018/NMT-4381 and by the HexaGB project of the Spanish Ministry of Science (reference RTI2018-098245). Computer resources and technical
assistance provided by the Centro de Supercomputaci\'on y Visualizaci\'on de Madrid
(CeSViMa) are gratefully acknowledged. Additionally, the authors thankfully
acknowledge the computer resources at CTE-Power and Minotauro in the Barcelona
Supercomputing Center (project QS-2021-1-0013). Finally, use of  the
computational resources of the Center for Nanoscale Materials, an Office of
Science user facility, supported by the U.S. Department of Energy, Office of
Science, Office of Basic Energy Sciences, under Project No. 73377, is
gratefully acknowledged. CMA also acknowledges the support from  the Spanish
Ministry  of  Education through the Fellowship FPU19/02031. 

\footnotesize


\providecommand{\url}[1]{\texttt{#1}}
\providecommand{\urlprefix}{}
\providecommand{\foreignlanguage}[2]{#2}
\providecommand{\Capitalize}[1]{\uppercase{#1}}
\providecommand{\capitalize}[1]{\expandafter\Capitalize#1}
\providecommand{\bibliographycite}[1]{\cite{#1}}
\providecommand{\bbland}{and}
\providecommand{\bblchap}{chap.}
\providecommand{\bblchapter}{chapter}
\providecommand{\bbletal}{et~al.}
\providecommand{\bbleditors}{editors}
\providecommand{\bbleds}{eds.}
\providecommand{\bbleditor}{editor}
\providecommand{\bbled}{ed.}
\providecommand{\bbledition}{edition}
\providecommand{\bbledn}{ed.}
\providecommand{\bbleidp}{page}
\providecommand{\bbleidpp}{pages}
\providecommand{\bblerratum}{erratum}
\providecommand{\bblin}{in}
\providecommand{\bblmthesis}{Master's thesis}
\providecommand{\bblno}{no.}
\providecommand{\bblnumber}{number}
\providecommand{\bblof}{of}
\providecommand{\bblpage}{page}
\providecommand{\bblpages}{pages}
\providecommand{\bblp}{p}
\providecommand{\bblphdthesis}{Ph.D. thesis}
\providecommand{\bblpp}{pp}
\providecommand{\bbltechrep}{Tech. Rep.}
\providecommand{\bbltechreport}{Technical Report}
\providecommand{\bblvolume}{volume}
\providecommand{\bblvol}{Vol.}
\providecommand{\bbljan}{January}
\providecommand{\bblfeb}{February}
\providecommand{\bblmar}{March}
\providecommand{\bblapr}{April}
\providecommand{\bblmay}{May}
\providecommand{\bbljun}{June}
\providecommand{\bbljul}{July}
\providecommand{\bblaug}{August}
\providecommand{\bblsep}{September}
\providecommand{\bbloct}{October}
\providecommand{\bblnov}{November}
\providecommand{\bbldec}{December}
\providecommand{\bblfirst}{First}
\providecommand{\bblfirsto}{1st}
\providecommand{\bblsecond}{Second}
\providecommand{\bblsecondo}{2nd}
\providecommand{\bblthird}{Third}
\providecommand{\bblthirdo}{3rd}
\providecommand{\bblfourth}{Fourth}
\providecommand{\bblfourtho}{4th}
\providecommand{\bblfifth}{Fifth}
\providecommand{\bblfiftho}{5th}
\providecommand{\bblst}{st}
\providecommand{\bblnd}{nd}
\providecommand{\bblrd}{rd}
\providecommand{\bblth}{th}

\end{document}